\documentclass[a4paper,10pt,twocolumn,aps]{revtex4}
\usepackage{isolatin1}
\usepackage{indentfirst,bbold,amsmath,amssymb,mathrsfs}
\usepackage[all]{xy}

\begin{document}

\title{Conserved gravitational charges, locality and the holographic Weyl anomaly - a fresh viewpoint}
\author{Pedro Lauridsen Ribeiro}
\affiliation{Departamento de Física Matemática -- Instituto de Física,
Universidade de São Paulo \\ CP 66.318\qquad 05315-970\qquad São Paulo, SP --- Brasil}
\email{pribeiro@fma.if.usp.br}
\thanks{This is a shortened version of a poster presentation made
at the International Conference ''100 Years of Relativity'', held 
at São Paulo, Brazil, in August 22nd-24th, 2005. This project is supported by 
FAPESP under grant no. 01/14360-1.}

\begin{abstract}
Since the proposal of the AdS/CFT correspondence, made by Maldacena and Witten, there has been some controversy about the definition of conserved Noether charges associated to asymptotic isometries in asymptotically AdS spacetimes, namely, whether they form an anomalous (i.e., a nontrivial central extension) representation of the Lie algebra of the conformal group in odd bulk dimensions or not. In the present work, we shall review the derivation of these charges by using covariant phase space techniques, emphasizing the principle of locality underlying it. We shall also comment on how these issues manifest themselves in the quantum setting.
\end{abstract}

\maketitle

\section{Introduction}

The issue of defining global conserved charges in General
Relativity is quite a delicate one. The difficulties one finds stem from
the fact that the dynamical problem posed by Einstein's equations 
is a rather \emph{sui generis} one: It does not possess a local 
dynamics in the usual sense of a ``time evolution'', because the
very choice of ``time'' is a local (gauge) symmetry of the
system (Hamiltonian constraint), and thus the Noether current
corresponding to it, which was supposed to generate the dynamics
according to the usual Hamiltonian recipe, actually vanishes
``on shell''. This stems from the deep physical fact, put forward
by Mach and Einstein, that one needs a material, physical
procedure to fix a local notion of time (namely, to fix a 
coordinate system), in a way that the physical laws ruling
these procedures are also local and moreover independent 
of this choice -- namely, one needs a nonvanishing matter
energy-momentum tensor. This is the physical content of Einstein's
equations. Another, distinct way of fixing the dynamical
interpretation (namely, the improvement terms to be added 
to the constraints of the theory) is to assume that the metric
approaches some fixed, symmetrical background near infinity
for which we do know how to fix such an interpretation.\\

However, one must be careful with formal computations 
regarding the on-shell action -- its value for the whole spacetime
may be infinite. This divergence inserts a new 
ambiguity in the definition of conserved gravitational charges, 
this one depending only on the conformal structure of infinity.
Whereas the usual definition gives a family of charges whose
Poisson algebra is a true representation of the algebra of asymptotic
isometries, the potential ambiguity stemming from the above
divergence may lead to an obstruction to forming a 
representation\cite{sken}. The inception of Maldacena-Witten AdS-CFT
correspondence, according to which the asymptotic boundary behaviour
of bulk fields corresponds to sources for ``dual'', 
boundary conformal fields, has led on its turn to another, QFT-inspired
prescription, according to which nontrivial (with respect to pure AdS)
boundary behaviour of the bulk metric acts as a source for the 
dual energy-momentum tensor, whose renormalization yields
a trace anomaly in even boundary dimensions.\\

Here, we start from the fact that the variational principle
determining dynamics is \emph{inherently local}. Namely, one 
must keep track of the supports of the functionals -- conserved 
gravitational charges are \emph{quasi-local} quantities, and
they must be obtained as suitable limits of local quantities. 
In this sense, conserved gravitational currents are indeed localized 
at infinity, since local gravitational charges, as we've seen, don't 
actually exist (more precisely, they vanish everywhere). The 
near-boundary behaviour ends up being affected by the same kind of 
ambiguity one faces in the renormalization of QFT\cite{scharf}. The 
result is, however, thatthe anomalous terms are actually independent 
of the detailed behaviour of the bulk metric (they depend only on the 
conformal class of the boundary metric)\cite{him}, and must be fixed 
by additional, physical considerations. The AdS-CFT recipe\cite{sken} 
is one possible answer. \\

We review the calculations leading to the above ambiguity in Section 2 
in an unified way by using the covariant phase space formalism (Peierls
bracket\cite{him}). In Section 3, we propose what we believe to be a 
QFT counterpart of the action of gravitational charges, \emph{based on 
first principles} -- namely, locality, covariance and causality. By 
employing a functorial framework in the spirit of \cite{bfv}, we give a 
prescription, inspired by our previous classical calculations, which seem 
to capture their essential aspects. 

\section{Covariant phase space approach to conserved charges}

Let $\mathscr{M}$ be a $n$-dimensional manifold, $K\Subset\mathscr{M}$
with regular boundary $\partial K$. Define 

\begin{equation}\label{p1}
S_K[g,\Lambda]=\int_K(R(g)+2\Lambda)\sqrt{-g}d^nx,
\end{equation}
($\text{Ric}(g)$ and $R(g)$ are resp. the Ricci tensor and the scalar
curvature associated with $g$) the Einstein-Hilbert action with 
cosmological constant $\Lambda$. For variations $\delta g$ supported 
in the interior of $K$, we have 

\begin{equation}\label{p2}
\delta S_K[g,\Lambda]=\int_K\underbrace{\left(\text{Ric}(g)-\frac{1}{2}R(g)g+
\Lambda g\right)}_{\doteq G^\Lambda_{ab}}\delta g\sqrt{-g}d^n x=0
\end{equation}

at a $g$ satisfying the Eisntein equations
$\text{Ric}(g)-1/2Rg+\Lambda g=0$. Now, consider an \emph{arbitrary} 
metric (i.e., not necessarily ``on shell''), and perform metric 
variations $\delta g^{ab}=1/2\nabla^{(a}X^{b)}$ coming from 
(infinitesimal) spacetime diffeomorphisms $X^a$. The Bianchi
identities imply that:

\begin{equation}\label{p3}
G^\Lambda_{ab}\nabla^aX^b=\nabla^b(G^\Lambda_{ab}X^a).
\end{equation}

If $X$ if timelike, one can see that $\mathscr{C}_a(g,X)\doteq 
G^\Lambda_{ab}X^b$ does not contain second derivatives of the 
metric in the $X$ direction. In other words, the $\mathscr{C}_a(g,X)$'s 
are \emph{constraints} of the theory, expressing its diffeomorphism 
invariance. Likewise, these constraints determine the dynamics of any 
gauge-invariant quantity -- instead of showing it in general, we'll 
specialize to asymptotically AdS (AAdS) spacetimes, for which the
``boundary improvement'' procedure mentioned in the Introduction 
can be implemented rather explicitly.\\

Let $(\widehat{\mathscr{M}},\widehat{g})$ be a $n$-dimensional
$(n\geq 4)$, AAdS sapcetime with conformal factor $\Omega$, conformal
completion $(\mathscr{M},g\doteq\Omega^2\widehat{g})$ and conformal 
boundary $(\mathscr{I}\doteq\partial\mathscr{M},q\doteq 
g\upharpoonright\!_{\mathscr{I}})$, with $q$ lying within the 
conformal class $[q]$ of the Einstein static universe (ESU).
We'll assume that our AAdS spacetimes satisfy empty space Einstein 
equations everywhere and are causally simple (i.e., the causal past 
and future of any compact set are closed), hence stably causal, as 
in \cite{ribeiro}. This means they can be foliated by equal-time 
surfaces (not necessarily Cauchy) by means of a global time function, 
say $\tau$. We can suppose that $\tau$ can be smoothly extended to 
$\mathscr{M}$ in such a way that $\tau\upharpoonright\!_{\mathscr{I}}$ 
is also a global time function.\\

We shall consider the following setting: Let $t_1<t_2\in\mathbb{R}$, 
$\epsilon\in[0,\zeta)$, $\zeta$ such that $d\Omega$ vanishes nowhere 
in the collar $\mathscr{I}\times[0,\zeta)$, and set $\Sigma_t=
\tau^{-1}(t)$, $\Sigma^\epsilon_t=\Sigma_t\setminus\Omega^{-1}([0,
\epsilon]$. The ingoing and outgoing null hypersurfaces emanating from
$\partial\Sigma^\epsilon_{t_i}$, $i=1,2$ cross resp. in $\Delta_i$,
$\Delta_o$, forming the edges of the Cauchy surfaces resp. for the regions
$\mathscr{O}$, $\mathscr{O}_1$. By choosing $t_1$ and $t_2$
sufficiently close to each other, we can assure that $\Delta_i$ and
$\Delta_o$ are smooth (in such a case, $\mathscr{O}$ and
$\mathscr{O}_1$ are regular diamonds) and belong to the same Cauchy
surface for $\mathscr{O}_1$. We also set $\Omega(\Delta_o)
\subset[\epsilon',\epsilon'']$, $0<\epsilon'<\epsilon''<\epsilon$.
Finally, we remark that we can deform the orbits of $\tau$ by suitably
redefining the latter, in a way that each orbit of 
$\Sigma_t\setminus\Sigma^\epsilon_t$ under $\tau$ belong to some
level surface of the collar above.\\

We'll compute all quantities we need from the Einstein-Hilbert
action $S_{K_{t_1,t_2,\epsilon}}[\widehat{g},\Lambda]$, where $K_{t_1,t_2,\epsilon}
=\cup_{t\in[t_1,t_2]}\Sigma^\epsilon_t\Subset\widehat{\mathscr{M}}$.
Given an arbitrary vector field $X^a$, the variation of the action
under $X^a$ ($\delta g^{ab}=1/2\nabla^{(a}X^{b)}$) reduces to
(we'll omit the volume elements for simplicity)

\[
\delta_X S_{K_{t_1,t_2,\epsilon}}[\widehat{g},\Lambda]=-\int^{t_2}_{t_1}\int_{\partial
\Sigma^\epsilon_t}\mbox{\large(}G^\Lambda_{ab}\omega^a X^b +\]
\begin{equation}
+\omega^a\theta(\widehat{g},\mathfrak{L}_X\widehat{g})_a\mbox{\large)}
+\int_{\Sigma^\epsilon_{t_2}-
\Sigma^\epsilon_{t_1}}\left(G^\Lambda_{ab}\sigma^a X^b+\sigma^a
\theta(\widehat{g},\mathfrak{L}_X\widehat{g})_a\right)\label{p4}
\end{equation}

($w^a$ and $\sigma^a$ are the unit ingoing spacelike, resp. 
future directed timelike normals to the timelike, resp. spacelike
smooth piecewise components of $\partial K_{t_1,t_2,\epsilon}$).
The 1-form $\theta(\widehat{g},\delta\widehat{g})$ is Poincaré dual (with respect to 
the volume element $\sqrt{-\widehat{g}}$) to the boundary term
coming from an arbitrary first variation $\delta\widehat{g}$. Consider now
the following, arbitrary second variation, imposing the following 
conditions on $X^a$:

\begin{itemize}
\item $X^a$ vanishes in a neighborhood $N_1$ of $\Sigma_{t_1}$;
\item On a neighborhood $N_2$ of $\Sigma_{t_2}$, disjoint from $N_1$,
$X^a$ is an \emph{asymptotic Killing field}, i.e., $\mathfrak{L}_X
(\widehat{g})$ decays at least as fast as $\epsilon^{n-2}$
as $\epsilon\rightarrow 0$ (this implies that the integral of 
the Lie derivative over $\partial\Sigma^\epsilon_t$ has a finite
limit as $\epsilon\rightarrow 0$).
\end{itemize}

This can be done by multiplying an arbitrary asymptotic Killing 
field $X^a$ by a regularized step function.
Now here comes the main point: if $\delta\widehat{g}$ is tangent to
a curve of solutions of the Einstein equations, it's propagated 
(in a particular gauge for the linearized equations) by convolution
with the covariant derivative along the orbits of $\tau$ of some 
fundamental solution $E_{cdef}$ associated with the globally
hyperbolic region $\mathscr{O}_1$, along the timelike component of 
the boundary (Duhamel's principle). Let us drop the variation of 
$\theta$ for now; we shall now exploit the presence of the 
step function and choose the \emph{retarded} fundamental solution.
Putting it all together
from (\ref{p4}), we get:

\begin{eqnarray}
-\delta\int^{t_2}_{t_1}\int_{\partial
\Sigma^\epsilon_t}G^\Lambda_{ab}\omega^a X^b & = & \int^{t_2}_{t_1}\int_{\partial
\Sigma^\epsilon_t}\omega^a X^b\partial_t\delta g^{ab}.\label{p6}
\end{eqnarray}

The calculation of the variation of $\theta$ is more cumbersome and 
we'll skip it due to lack of space\cite{him}.
Antisymmetrization of the second variation and repetition of the
procedure above leads to the usual expression for the charges,
which is no longer dependent neither on the regularized step function, nor on $t$:

\begin{equation}\label{p7}
Q(X)\varpropto 
\lim_{\epsilon\rightarrow 0}\int_{\partial\Sigma^\epsilon_t}
\Omega^{3-n}C_{abcd}(g)\nabla^a\Omega\nabla^b\Omega X^c\sigma^d,
\end{equation}

where $C_{abcd}(g)$ is the Weyl tensor of $g$. Notice that 
the limit is taken by simultaneously taking $t_1 \rightarrow t_2$ 
and shrinking the regularized step function to the Heaviside
step function, so that $\mathscr{O}_1$ remains globally 
hyperbolic all the way -- otherwise, the retarded fundamental 
solution may cease to exist or no longer be unique. The requirement 
that the variation of the charges should act 
as boundary sources for linearized gravity around a solution of
Einstein's equations fix the charges themselves up to dynamically
trivial terms. More precisely, it fixes 
their Peierls bracket with compactly supported, gauge invariant 
functionals of the on shell metric (local observables). The remaining 
ambiguity has vanishing Peierls bracket with all
local observables and depends on the scaling
properties of the local charges w.r.t. $\Omega$ as the limit is taken,
as it arises from the ill-defined multiplication of the limit retarded
fundamental solution by a Heaviside step function.
Depending on the asymptotic scaling degree\cite{scharf}, the 
needed extension to $\mathscr{I}$ defined by fixing the ambiguity
may acquire unavoidable logarithmic terms, violating the expected
scaling behaviour. This happens, for instance, for $n$ odd\cite{sken}.

\section{The picture from Local Quantum Physics}

The conceptual advantage of employing the Peierls bracket in the classical
calculations above is that it brings the principle of locality to the
forefront, in a way akin to QFT. One can emulate the line of reasoning
above within local quantum physics (algebraic QFT\cite{haag})
by means of the functorial formalism proposed in \cite{bfv}.
Let $\underline{\mathscr{M}an}$ is the category of strongly causal, $n$-dimensional
spacetimes $(\mathscr{O},g)$, with arrows defined by orientation-preserving isometric embeddings with open, 
causally convex images, and $\underline{\mathscr{A}lg}$ the
category of unital C*-algebras, whose arrows are unit-preserving C*-morphisms.
A \emph{locally covariant quantum theory} is simply a \emph{covariant functor}
$\mathfrak{A}$ between both categories, i.e., the diagram

\begin{equation}\label{p8}
\xymatrix{(\widehat{\mathscr{M}},\widehat{}g)\ar[d]_{\mathfrak{A}}\ar[r]^\psi &
(\widehat{\mathscr{M}'},\widehat{g'}) \ar[d]_{\mathfrak{A}}\ar[r]^\psi' &
(\widehat{\mathscr{M}''},\widehat{g''}) \ar[d]_{\mathfrak{A}} \nonumber\\
\mathfrak{A}(\widehat{\mathscr{M}},\widehat{g})\ar[r]^{\mathfrak{A}\psi} & 
\mathfrak{A}(\widehat{\mathscr{M}'},\widehat{g'})\ar[r]^{\mathfrak{A}\psi'} & 
\mathfrak{A}(\widehat{\mathscr{M}''},\widehat{g''})}
\end{equation}

\vspace*{0.3cm}

commutes. We say that $\mathfrak{A}$ is
\emph{causal} if the local algebras at spacelike separated regions
commute, and \emph{primitively causal} if the embedding of any
neighborhood of a Cauchy surface into its Cauchy development induces
an isomorphism of the respective algebras. In the latter case, 
one can define retarded and advanced ``scattering morphisms'' by
suitable metric perturbations. The composition of both gives an
automorphism $\beta{g}$ of $\mathfrak{A}(\widehat{\mathscr{M}},\widehat{g})$
(relative Cauchy evolution) whose functional derivative

\begin{equation}\label{p9}
\frac{i}{2}\langle\Phi,[T^{\mu\nu}(x),\pi_\omega(A)]\Phi\rangle\doteq
\frac{\delta}{\delta g}\langle\Phi,\pi_\omega(\beta_{g}(A))\Phi\rangle
\end{equation}

acts in the same way (i.e., as a densely defined derivation) as 
the commutator with the energy-momentum tensor, in the sense of 
quadratic forms, on the GNS Hilbert space induced by a state 
satisfying the microlocal spectrum condition,
endowed with boundary conditions in the following sense: it approaches an
``Rindler-Unruh'' type equilibrium state after a sufficiently long
relaxation time -- for such states, it's meaningful to speak about the
implementation of asymptotic isometries. This derivation on the
local algebras shares many properties of the Peierls bracket, and
the splitting of the second variation into their retarded and advanced
parts is also subject to renormalization ambiguities, in the same
sense as above. 

\begin{acknowledgments}
I'd like to thank prof. Michael Forger for the enlightening discussions
on the principle of locality in classical field theory, which led to 
the crucial physical insights at the basis of the present work.
\end{acknowledgments}

\end{document}